\documentclass{article}

\usepackage[active]{srcltx}

\usepackage{amssymb}
\usepackage{amsmath}
\usepackage[dvips]{graphicx}
\usepackage{setspace}
 \usepackage[left=4cm,right=4cm,top=4cm,bottom=4cm]{geometry}

\def\func#1{\mathop{\rm #1}}

\renewcommand{\thefootnote}{\fnsymbol{footnote}}

\begin{document}

{\centerline{\textbf{\large Temperature profile in a liquid-vapor
interface}}} \vskip 0.1cm {\centerline{\textbf{\large near the critical point}}}\vskip 0.2cm

{\centerline{Henri Gouin\,\footnote {Aix-Marseille Univ,
Centrale Marseille, CNRS, M2P2 UMR 7340, 13451 Marseille France\\
E-mail: henri.gouin@univ-amu.fr, henri.gouin@yahoo.fr}\quad
\&\quad Pierre Seppecher\,\footnote{IMATH, Institut de Math\'ematiques,
 Universit\'e de Toulon,
 BP 132, 83957 La Garde, Cedex, France \\ E-mail: seppecher@imath.fr}
\footnote{ LMA, CNRS, UPR 7051, Aix-Marseille Univ, Centrale Marseille,  13402 Marseille, France\\ E-mail: seppecher@lma.cnrs-mrs.fr}}}

\vskip 0.5cm
 {\textbf{Abstract  }}

\medskip   Thanks to an expansion   with respect to densities of
energy, mass and entropy, we discuss the concept of
\emph{thermocapillary fluid} for inhomogeneous fluids.  The
non-convex state law valid for homogeneous fluids is modified by
adding terms taking into account the gradients of these densities.
This seems more realistic than Cahn and Hilliard's model which
uses a density expansion in mass-density gradient  only. Indeed,
through liquid-vapor interfaces, realistic potentials in molecular
theories show that
   entropy density and temperature do  not vary  with the mass density as
   it would do in  bulk phases.

In this paper we prove, using a rescaling process near the
critical point,
 that  liquid-vapor interfaces behave essentially in
the same way as   in Cahn and Hilliard's model.

\bigskip



\medskip

\renewcommand{\thefootnote}{\arabic{footnote}}

\section{Introduction}

Phase separation between liquid and vapor is due to the fact that
density of internal energy (i.e. internal energy per unit volume) $\varepsilon_{0} (\rho ,\eta)$ of
homogeneous fluids is a non-convex function of  mass density $\rho$ and  entropy density $\eta$. At a given temperature $T_0$, this
non-convexity property is related with the non-monotony of
thermodynamical pressure $P(\rho,T_0)$.

The reader
may be accustomed to use specific quantities $\alpha={\varepsilon}/{\rho}$, $s={\eta}/{\rho}$ and $v=1/\rho$ instead of volume densities. Indeed the non-convexity property of $\varepsilon_{0}$ is equivalent to the non-convexity of  $\alpha$ as a function of $s$ and $v$. In this paper, in accordance with Cahn-Hilliard standard presentation, we privilege volume densities.

In continuum  mechanics the simplest model for describing
inhomogeneous fluids inside interfacial layers considers an
internal-energy density $\varepsilon$ as the sum of two terms: the first
one  previously defined as $\varepsilon_{0}(\rho ,\eta)$,
corresponds to the fluid with an uniform composition equal to its
local one, and the second one  associated with the non-uniformity
of the fluid is approximated by a gradient expansion,
\begin{equation}
\varepsilon :=\varepsilon_{0} (\rho ,\eta)+{\frac{1}{2}}%
\,m\,|\func{grad}\rho\,|^{2}, \label{cahnenergy}
\end{equation}
where $m$ is a coefficient assumed to be independent of $\rho$,
$\eta$ and $\func{grad}\rho $. This form of internal energy density
can be deduced from molecular mean-field theory where the
molecules are modeled as hard spheres
submitted to Lennard-Jones potentials \cite{Widom,GouinJCP}.

This energy has been introduced by  van der Waals \cite{Waals} and
is widely used in the literature
\cite{Korteweg,Ono,Cahn,casal,casal1}.  This model, nowadays
  known as Cahn-Hilliard  fluid model, describes interfaces
  as diffuse layers. The
mass density profile connecting liquid
to vapor becomes a smooth  function.

The model has been widely used   for describing micro-droplets
 \cite{Isola2,Isola3},  contact-lines
\cite{casal2,seppecher,Seppecher1,Gouin1}, nanofluidics
\cite{Gouin2,Gouin9,Garajeu}, thin films \cite{Gouin6},
  vegetal biology \cite{gouin7,gouinbio}.
It has been extended to more complex situations e.g.
 in
  fluid mixtures, porous materials\dots, thanks
to the so-called  second-gradient theory \cite{Germain,Isola}
which models the behavior of strongly inhomogeneous media
\cite{Gouin-Ruggeri,Ruggeri,dell'Isola,Eremeyev,Dell'Isola1,Forest}.

It has been noticed that,  at equilibrium, expression
\eqref{cahnenergy} for the energy density
 yields an uniform temperature $T_{_0} $ everywhere in
  inhomogeneous fluids,
\begin{equation}
T:=\frac{\partial \varepsilon_{_0} }{\partial \eta}(\rho
,\eta)=T_{_0} . \label{temp0}
\end{equation}
Let us note that it is not the same for   chemical potential
\begin{equation*}
\mu_{_0}:= \frac{\partial \varepsilon_{_0} }{\partial \rho}(\rho
,\eta) ,
\end{equation*}
which takes the same values in the different bulks but is not
uniform inside interfacial layers. From Eq. \eqref{temp0} one can
deduce that
 the entropy density varies with the
mass density   in the same way as in the bulks and it is a
peculiarity of the Cahn-Hilliard model  that the
configurational $\eta$ and $\varepsilon$ can be written in term of
$\rho$, only. The points $(\rho , \eta, \varepsilon)$ representing
phase states   lie on  curve $T=T_0$  and such a model  inevitably
lead to monotonic variations of all densities \cite{Widom}.
Original assumption \eqref{cahnenergy} of van der Waals which uses
long-ranged but weak attractive
  forces   is not exact for more realistic intermolecular potentials
\cite{Ornstein,Hamak,Evans}.   Aside from the
question of accuracy, there are qualitative features like
non-monotonic behaviors in transition layers, especially in
systems of more than one component, that require two or more
independently varying densities - entropy included - (see chapter
3 of \cite{Rowlinson}).
For these reasons,  model \eqref{cahnenergy} has been extended in
\cite{Rowlinson,casal4} by taking into account  not only the
strong mass density variations through interfacial layers but also
  the strong variations of entropy associated with latent-heat of
phase changes.  Rowlinson and Widom in \cite{Rowlinson} (chapter 3 and
chapter 9) noticed that
  $T=T_{_0}$ is not exact through liquid-vapor interfaces and they
 introduced an energy arising from the mean-field theory and
depending on densities $\rho$ and $\eta $ and also on the
gradients of these densities; furthermore, they said that
\emph{near the critical point, a gradient expansion typically
truncated in second order, is most likely to be successful and
perhaps even quantitatively accurate}. This extension  has been
called \emph{thermocapillary fluid model} in
\cite{casal4} and used in different  physical situations when the
temperature varies  in strongly inhomogeneous parts of complex
media \cite{casal4,Gouin,Forest1,Forest2,Maitournam}.

Near a single-fluid critical point,   the mean-field molecular
theory yields an approximate
 but realistic behavior  \cite{Rowlinson,Domb}.  In mean-field
theory, the differences of thermodynamical quantities between liquid and vapor phases are expressed
 in power laws of the difference between   temperature  and   critical  temperature. Transformations from liquid to vapor are associated with
second-order phase transitions  and the mass density difference
between the two phases  goes to zero as the temperature is
converging to the critical one. The same phenomenon  holds true
for the latent-heat of phase transition and for the difference of
 entropy densities between liquid and vapor phases.

 In this paper we neglect gravity and we use a slightly
more general model. We consider state laws  which link
 densities $\varepsilon,  \rho ,  \eta $  and their
gradients. We  derive the liquid-vapor equilibrium equations of
non-homogeneous fluids. As,  at equilibrium, a given total mass of
the fluid in a fixed domain   maximizes its total entropy
while its total energy remains constant, the problem can be studied in a variational framework.

We make explicit a polynomial expansion of the homogeneous state
law near the critical point. In convenient units, we obtain a
generic expression depending only on a unique parameter $\chi$.

We introduce a small parameter $\kappa$ which measures the distance
of the considered equilibrium state to the critical point. Using a
rescaling process near
the critical point we obtain mass and temperature profiles
 through the liquid-vapor interface. The
magnitude orders with respect to $\kappa$ of mass, entropy,
temperature are analyzed. The variations of temperature and
entropy density inside the interfacial layer appear  to be of an order less
than the variation of mass density. Consequently, neglecting these
variations is well-founded and justifies the utilization of Cahn-Hilliard's model  near the critical point and indeed we prove that the mass density profile
converges towards the classical profile obtained by using the Cahn-Hilliard model which does
not take account of variations of entropy  density. A conclusion
highlights these facts.

\section{\label{sec2}Equations of equilibrium}

\subsection{\label{subsec2.1}Preliminaries}

When homogeneous simple fluids are considered, a state law
$$
{\cal L}_0(\varepsilon,\eta,\rho)=0
$$
links internal
energy  density $\varepsilon$, entropy  density $\eta$ and
mass density $\rho$. This local law is generally made explicit under the
form
$$
\varepsilon-\varepsilon_{_0}(\eta,\rho)=0 .
$$
In other words, it is   assumed without loss of generality that
$\partial  {\cal L}_0/{\partial \varepsilon}=1$. Then, as usual,
one introduces the Kelvin temperature $T:=-\partial {\cal
L}_0/{\partial \eta}$, the chemical potential ${\mu }:=-\partial
{\cal L}_0/{\partial \rho}$ and the thermodynamical pressure
$$
P:=\rho\,{\mu } -\varepsilon+\eta\,T.
$$
These notations can be resumed as
\begin{equation*}
d\varepsilon- {\mu }\, d\rho-{ T}\, d\eta=0, \quad dP- \eta\,dT
-\rho\, d\mu=0. \label{chempot}
\end{equation*}

However, when the state of the material endows strong spatial
 variations of the thermodynamical variables -
 as it is the case  near a liquid vapor interface -
    the locality of the state law has to be questioned.
    This is what we do in this paper by considering
    a general law of the type
  \begin{equation}
    {\cal L}(\varepsilon,\eta,\rho,\nabla \varepsilon,
    \nabla \eta,\nabla \rho )=0 ,\label{relation}
  \end{equation}
    where $\nabla$ denotes the spatial gradient.   For the sake of simplicity, we study in this paper
      the   particular form (\footnote[1]{Let us note that the case
      $\varepsilon -
      \varepsilon_0(\eta,\rho) -\frac{1}{2}
  \Big(
  C_0\, |\nabla\rho|^2  + 2 D_0\,
  \nabla\rho\cdot\nabla\eta + E_0\, |\nabla\eta|^2
  \Big) = 0 $ has been considered in \cite{Rowlinson}, chapter 3.})
      :
\begin{align}
  {\cal L}(\varepsilon,\eta,\rho,\nabla \varepsilon,\nabla \eta,\nabla \rho )= &
   {\cal L}_0(\varepsilon,\eta,\rho)    -  \frac{1}{2}
  \Big(
  C_0\, |\nabla\rho|^2 + E_0\, |\nabla\eta|^2 + H_0\, |\nabla\varepsilon|^2\nonumber \\
  + & 2\, D_0\, \nabla\rho\cdot\nabla\eta + 2\, F_0\, \nabla\rho \cdot \nabla\varepsilon + 2\, G_0 \, \nabla\eta \cdot \nabla \varepsilon
  \Big),
  \label{specialvolumeenergy}
\end{align}
where $$ \left[\begin{matrix}C_0 & D_0 & F_0 \cr D_0 & E_0 & G_0
\cr F_0 & G_0 & H_0 \end{matrix}\right]
$$ is   a
constant positive symmetric matrix. This is the
simplest extension of the classical model when one wants to take
account of the spatial variations of $\eta$, $\varepsilon$ and
$\rho$. Generalization \eqref{relation} is widely studied \cite{Waals,Cahn} in the particular case  ${\cal L}(\varepsilon,\eta,\rho,\nabla \rho )=0$; that is when one sets  $D_0=E_0=F_0=G_0=H_0=0$ in \eqref{specialvolumeenergy}. This special case coincides with
the well-known model of Cahn-Hilliard's fluids \cite{Cahn}.

In
our framework, we still call temperature, chemical potential,
thermodynamical pressure  the quantities
$$
T:=-\partial {\cal L}_0/{\partial \eta},\quad {\mu }:=-\partial
{\cal L}_0/{\partial \rho}\quad {\rm and}\quad P:=\rho\,{\mu }
-\varepsilon+\eta\,T.
$$
 Thus,  the state law
reads in differential form :
\begin{equation}
d\varepsilon- \mu\,d\rho    - {T}\, d\eta - \boldsymbol{\Phi}\cdot
d(\nabla \rho) - \boldsymbol{\Psi}\cdot d(\nabla\eta) -
\boldsymbol{\Xi}\cdot d(\nabla\varepsilon) =0 \label{diffenergy}
 \end{equation}
with
\begin{equation*}
  \begin{split}
\boldsymbol{\Phi} &= C_0\, \nabla\rho + D_0 \, \nabla\eta+ F_0 \, \nabla\varepsilon,\quad \\
\boldsymbol{\Psi} &= D_0\, \nabla\rho + E_0\, \nabla\eta +  G_0 \, \nabla\varepsilon,\quad \\
\boldsymbol{\Xi} &= F_0\, \nabla\rho + G_0\, \nabla\eta +  H_0 \,
\nabla\varepsilon.
  \end{split}
  \label{phipsi}
 \end{equation*}

\subsection{\label{subsec2.2} The variational method}

The total mass and the total energy of an isolated and fixed
domain ${\cal D}$ are
\begin{equation*}
  M= \int_{\cal D}\rho \ dx,\qquad E = \int_{\cal D}\varepsilon \ dx,
\end{equation*}
where $dx$ is the volume element. They remain constant during the evolution of the system towards
equilibrium. The equilibrium is reached when the total entropy
$$ S=\int_{{\cal D}}\eta \,dx=\int_{{\cal D}}\rho\, s \,dx $$
of the system is maximal. With classical notations, at equilibrium
we get the variational equation
$$\delta S - T_0^{-1} (\delta E - \mu_0\, \delta M)=0$$
where $T_0^{-1}$ and $\mu_0$ are constant Lagrange multipliers
($T_0$ has the physical dimension of a temperature while $\mu_0$
has the physical dimension of a chemical potential). This equation
is valid for all variations $(\delta\varepsilon, \delta\eta,
\delta\rho)$ compatible with the state law i.e.
$\delta{\cal L}=0$. We can take this constraint into account by
introducing a Lagrange multiplier field $\Lambda$ (with
no physical dimension) and write that
$$T_0 \delta S - \delta E + \mu_0\, \delta M + \int_{{\cal D}}\Lambda\, \delta{\cal L}\, dx =0$$
for all triple field  $(\delta\varepsilon, \delta\eta,
\delta\rho)$. This equation reads
\begin{align*}
  \int_{\cal D} \Big( (T_0-\Lambda\, T)\,\delta \eta&+  (\Lambda- 1)\, \delta\varepsilon +
 (\mu_0-\Lambda\,\mu)\,\delta\rho \\
& - \Lambda\,\big( \boldsymbol{\Phi}\cdot (\nabla \delta\rho) +
\boldsymbol{\Psi}\cdot (\nabla\delta\eta) + \boldsymbol{\Xi}\cdot
(\nabla\delta\varepsilon) \big) \Big)\, dx=0
\end{align*}
Using the divergence theorem and considering only variations with
compact support in ${\cal D}$, we have
\begin{align*}
  \int_{\cal D} \Big((T_0- \Lambda\, T+\,\func{div}(\Lambda\boldsymbol{\Psi}) )\,\delta \eta&+(\Lambda-1+
  \func{div}(\Lambda\boldsymbol{\Xi}) ) \delta\varepsilon \\
  & + (\mu_0-\Lambda\,\mu+ \func{div}(\Lambda\boldsymbol{\Phi}))\,\delta\rho \Big)\,
  dx=0,
\end{align*}
and we deduce the local equations in ${\cal D}$ :
\begin{equation*}
  \begin{split}
    & \func{div}(\Lambda\boldsymbol{\Phi})=\Lambda\mu- \mu_0,\\
    &\func{div}(\Lambda\boldsymbol{\Psi}) =\Lambda T-T_0,\\
    &\func{div}(\Lambda\boldsymbol{\Xi})=1-\Lambda.
    \end{split}
  \end{equation*}
In the special case of a energy density of  form  \eqref
{specialvolumeenergy}, the system reads
\begin{equation}
\left\{
\begin{array}{l}
\displaystyle  C_0\, \func{div}(\Lambda\nabla\rho) + D_0\,\func{div}(\Lambda\nabla\eta) +F_0\, \func{div}(\Lambda\nabla\varepsilon) =\Lambda\mu- \mu_{_0},    \\
\displaystyle D_0\,\func{div}(\Lambda\nabla\rho) + E_0\,\func{div}(\Lambda\nabla\eta) +G_0\, \func{div}(\Lambda\nabla\varepsilon)= \Lambda T-T_0 , \\
\displaystyle F_0\,\, \func{div}(\Lambda\nabla\rho) + G_0\,\func{div}(\Lambda\nabla\eta) +H_0\, \func{div}(\Lambda\nabla\varepsilon)= 1- \Lambda , \label{€qphases}
\end{array}%
\right.
\end{equation}

\section{\label{sec3}Thermodynamical potentials near a critical point}

Let $(\varepsilon_c,\eta_c,\rho_c)$ be an admissible homogeneous
state indexed by $c$. Then,
$$
{\cal L}_0(\varepsilon_c,\eta_c,\rho_c)=0.
$$
 Let $P_c$, $T_c$, $\mu_c$
be the associated thermodynamical quantities.
At point $(\varepsilon_c,\eta_c,\rho_c)$, we assume that
$\partial^2{\cal L}_0/\partial \eta^2\not=0$  and we introduce the
quantity
$$
a_c:= \frac{\partial^2{\cal L}_0/\partial \eta\partial \rho}
{\partial^2{\cal L}_0/\partial \eta^2}(\varepsilon_c,\eta_c,\rho_c).
$$
If the studied fields remain close to point
$(\varepsilon_c,\eta_c,\rho_c)$, it is natural to make a change of
variables in order to work in the vicinity of zero; we set
\begin{align}
  &\tilde \rho:=\rho-\rho_c,\ \tilde \eta:=\eta-\eta_c + a_c \tilde \rho, \  \tilde \varepsilon:=\varepsilon- \varepsilon_c- (\mu_c-T_c a_c) \tilde \rho - T_c \tilde \eta,\label{change1}\\
  &\tilde{\cal L}_0(\tilde  \varepsilon,\tilde \eta,\tilde \rho):={\cal L}_0( \varepsilon_c
  +\tilde \varepsilon+T_c  \tilde \eta+(\mu_c-T_c a_c)\tilde \rho, \eta_c+ \tilde \eta-a_c \tilde \rho, \rho_c+\tilde\rho).\label{change2}
\end{align}
The change of variables \eqref{change1}-\eqref{change2} may seem unnecessarily complicated : its aim is, like in classical nondimensionalization process, to reduce the number of parameters of the problem to the minimal set of parameters which actually affect the qualitative features of the solution. We show below that a unique dimensionless parameter $\tilde \chi$ is enough for describing the shape of the energy function in the vicinity of the critical point.

It is clear that maximizing $\int_{\cal D} \eta\, dx$ under the
constraints $\int_{\cal D} \rho\, dx=M$ and  $\int_{\cal D}
\varepsilon\, dx=E$ is equivalent to maximizing  $\int_{\cal D}
\tilde \eta\, dx$ under the constraints $\int_{\cal D} \tilde
\rho\, dx=M-\rho_c |{\cal D}| $ and  $\int_{\cal D} \tilde
\varepsilon\, dx=E-\mu_c M -(\varepsilon_c+\mu_c\rho_c) |{\cal D}|
$. Therefore the variational analysis performed in the previous
section remains unchanged if we replace all quantities by their
 $ \widetilde{\ }\,$- equivalent. Of course this property is only true if we
replace the derivative quantities $T$, $\mu$ by the quantities
derived from $\tilde{\cal L}$. We  set:
\begin{equation}
  \tilde T:= T-T_c,\quad \tilde \mu:= \mu-\mu_c- a_c \tilde T \label{tempchem}
\end{equation}
The constants $(C_0,\dots ,H_0)$ have also to be modified but it is
not worth  writing the expressions of the new constants  $(\tilde
C_0,\dots ,\tilde H_0)$ in terms of $(C_0,\dots ,H_0)$,
$\varepsilon_c$, $\rho_c$, $T_c$, $\mu_c$ and $a_c$. We have
$\tilde {\cal L}_0(0,0,0)=0$, $ {\partial \tilde {\cal
L}_0}(0,0,0)/{\partial \tilde\eta}=0$, $ {\partial \tilde {\cal
L}_0}(0,0,0)/{\partial \tilde\rho}= 0 $ and, owing to the
particular choice we made by introducing $a_c$ in the change of variables, we have also
\begin{equation}
\frac{\partial^2 \tilde {\cal L}_0}{\partial
\tilde\eta\partial \tilde\rho}(0,0,0)=0 \label{partialder}.
\end{equation}
Consequently, from \eqref{tempchem} and \eqref{partialder}, we can write the Taylor expansion of $\tilde{\cal
L}_0$ in the vicinity of point $(0,0,0)$ under the form
\begin{align*}
  \tilde{\cal L}_0=\tilde \varepsilon -a_{20}\, \tilde \eta^2-  a_{02}\,\tilde  \rho^2 -
  a_{30}\, \tilde \eta^3 -  a_{21}\, \tilde \eta^2\tilde \rho -  a_{12}\, \tilde \eta\tilde \rho^2
  -  a_{03}\, \tilde \rho^3 +o(\tilde\tau^3)
\end{align*}
where $\tilde \tau$, which stands for $\max(\tilde \eta,\tilde  \rho)$, is a measure of the distance to point  $(\eta_c,\rho_c)$ in the space $(\eta,\rho)$. Indeed $
\tilde \tau \leq (1+|a_c|)\max(\eta-\eta_c,\rho-\rho_c)$.
Accordingly, we obtain:
$$
\tilde T=  2 a_{20} \tilde \eta +o(\tilde \tau).
$$
Recalling that we have assumed that $a_{20}=\partial^2\tilde {\cal
L}_0/\partial \tilde \eta^2\not=0$, we have $\tilde \tau \sim \max(\tilde
T,\tilde \rho)$ and $\tilde \eta=  {\tilde T}/{(2 a_{20})}
+o(\tau)$. Hence
\begin{align*}
  \tilde \mu&=  2 a_{02}\, \tilde \rho +  a_{21}\, \tilde \eta^2 + 2 a_{12}\, \tilde \eta\tilde \rho +
   3  a_{03}\,\tilde  \rho^2 + o(\tilde \tau^2)  \\
  &=  2 a_{02}\, \tilde \rho +  \frac{ a_{21}}{ 4 a_{20}^2}\, \tilde T^2 +
    \frac{ a_{12}}{  a_{20}}\,\, \tilde T\tilde \rho+ 3  a_{03}\,\tilde  \rho^2 +o(\tilde \tau^2).
\end{align*}

Now, we assume that $(\varepsilon_c,\eta_c,\rho_c)$ corresponds to
the critical point  of ${\cal L}_0$. Equivalently, $(0,0,0)$
is the critical point of $\tilde {\cal L}_0$.\newline The
critical conditions state that, at  {\it fixed} critical
temperature $\tilde T=0$,  the first and second derivatives of
$\tilde \mu$ with respect to $\tilde \rho$ vanish. In view of the
previous equation these conditions state that $a_{02}=a_{03}=0$.  Let us now go a bit further in the expansions of $\tilde L$, $\tilde T$ and $\tilde \mu$.
  In the generic case, when the coefficients $ a_{12}$ and $a_{04}$
like $a_{20}$ do not vanish, we get
\begin{align*}
  &\tilde{\cal L}_0=\tilde \varepsilon - a_{20} \tilde \eta^2   -  a_{12} \tilde \eta\tilde \rho^2 -  a_{04} \tilde \rho^4 +o( \tilde \xi^2), \\
  &\tilde T=  2 a_{20} \tilde \eta +  a_{12} \tilde \rho^2 +o( \tilde \xi),\\
  &\tilde \mu=  2 a_{12} \tilde \eta\tilde \rho+  4  a_{04} \tilde \rho^3+\tilde \rho\  o(   \tilde \xi),
\end{align*}
where $\tilde\xi$ stands for $\max\, (\tilde \eta, \tilde
\rho^2)$. Furthermore, we can use a mass unit such that $a_{04}=1$
and an entropy unit such that $a_{12}=1$. We denote $\tilde\chi$
the value of $a_{20}$ in such an unit system. We finally get
\begin{align*}
  &\tilde{\cal L}_0=\tilde \varepsilon - \tilde\chi \tilde \eta^2   -
    \tilde \eta\tilde \rho^2 -   \tilde \rho^4 +o(\tilde\xi^2) \nonumber\\
  &\tilde T=  2 \tilde\chi \tilde \eta +  \tilde \rho^2 +o(\tilde\xi)\\
  &\tilde \mu=  2 \tilde \eta\tilde \rho+  4 \tilde \rho^3+\tilde\rho \ o(\tilde\xi)\nonumber
\end{align*}
These equations are the generic asymptotic form of the
thermodynamic potentials near a critical point in an adapted
system of coordinates. Note that $\tilde \chi$ has to satisfy
$4\tilde\chi-1 > 0$ in order to ensure the positivity of
$\tilde\chi \tilde \eta^2 + \tilde \eta\tilde \rho^2 +   \tilde
\rho^4$. Otherwise no homogeneous phase could be stable in the
studied zone.

\begin{figure}[h]
\begin{center}
\includegraphics[width=10cm]{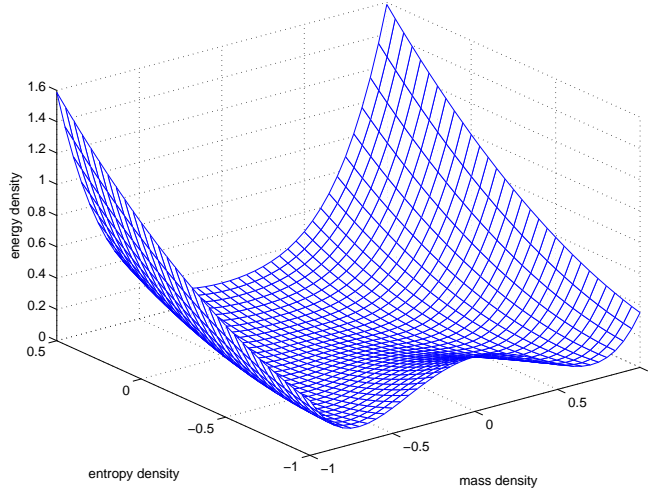}
\end{center}
\caption {Internal energy  density  $\tilde \varepsilon_0(\tilde \eta,\tilde \rho)$ of a homogeneous fluid near critical point
corresponding to
$(\tilde\eta_c,\tilde\rho_c,\tilde\varepsilon_c)=(0,0,0)$ when we
chose $\tilde\chi=0.35$\,. }   \label{fig1}
\end{figure}

From now on, we study the equilibrium of two phases by assuming that
\begin{align}
  &\tilde{\cal L}_0=\tilde \varepsilon - \tilde\chi \tilde \eta^2   -  \tilde \eta\tilde \rho^2 -  \tilde \rho^4 \label{Lzero}
\end{align}
and consequently
\begin{align}
  &\tilde T=  2 \tilde\chi \tilde \eta +  \tilde \rho^2, \label{temperature}\\
  &\tilde \mu=  2 \tilde \eta\tilde \rho+  4 \tilde \rho^3.\label{chempote}
\end{align}
Relations \eqref{temperature} and \eqref{chempote} are the associated temperature and chemical potential.
Function
\begin{equation}
 \tilde \varepsilon_0(\tilde \eta, \tilde \rho)=
\tilde\chi \tilde \eta^2   +  \tilde \eta\tilde \rho^2 +  \tilde
\rho^4 \label{energyinterne}
\end{equation}
is represented in Fig. \ref{fig1} where one can check that the
critical point lies on the boundary of the domain where
$\tilde\varepsilon $ does not coincide with its lower convex
envelope. 
\section{Integration of equations  in    planar interfaces}
We consider a planar interface and assume that all fields depend
only on   transverse space-variable $z$. We denote $\varphi'$ the
derivative of any field $\varphi$ with respect to $z$.

\subsection{System of equilibrium equations}
 {{System of equilibrium equations (\ref{€qphases}) completed by the state law reads in term of new  $ \widetilde{\ }\,$- equivalent quantities,
\begin{equation}
\left\{
\begin{array}{l}
 \tilde C_0\, (\tilde\Lambda \tilde \rho')' + \tilde D_0\,(\tilde\Lambda \tilde
 \eta')'+\tilde F_0\, (\tilde\Lambda \tilde \varepsilon')' =\tilde\Lambda \tilde \mu- \tilde \mu_{_0},    \\
 \tilde D_0\,(\tilde\Lambda\tilde  \rho')' + \tilde E_0\,(\tilde\Lambda \tilde
\eta')' +\tilde G_0\, (\tilde\Lambda \tilde \varepsilon')'=\tilde\Lambda
 \tilde  T-   \tilde T_0, \\
 \tilde F_0\, (\tilde\Lambda\tilde \rho')' + \tilde G_0\,(\tilde\Lambda\tilde
\eta')' +\tilde H_0\, (\tilde\Lambda \tilde \varepsilon')'= 1-  \tilde
\Lambda, \\
   \tilde{\cal L}_0(\tilde  \varepsilon,\tilde \eta,\tilde \rho) - \tilde Q( \tilde \rho',\tilde \eta', \tilde \varepsilon') = 0, \label{systemA}
\end{array}%
\right.
\end{equation}}}
where $\tilde Q( \tilde \rho',\tilde \eta', \tilde \varepsilon'):= \frac{1}{2}
  \big(\tilde C_0\, \tilde \rho'^2 + \tilde E_0\,\tilde \eta'^2
  +\tilde H_0\, \tilde \varepsilon'^2 +
2  \tilde D_0\,\tilde
\eta'\tilde \rho'
  + 2 \tilde F_0\, \tilde \rho'\tilde \varepsilon'+2 \tilde G_0\, \tilde \varepsilon'\tilde \eta'\big) $.

  Multiplying the three first equations respectively by $\tilde \rho'$, $\tilde \eta'$, $\tilde \varepsilon'$, summing and using the fourth equation derived with respect to $z$, leads to
  $$2 \Big(\tilde\Lambda\,  \tilde Q( \tilde \rho',\tilde \eta', \tilde \varepsilon')\Big)'=\tilde \varepsilon'-\tilde \mu_{_0}\tilde \rho'- \tilde T_0 \tilde
  \eta' ,$$
  which gives the first energy integral
  \begin{equation}
   2 \tilde\Lambda\,  \tilde Q( \tilde \rho',\tilde \eta', \tilde \varepsilon')=\tilde \varepsilon-\tilde \mu_{_0}\tilde \rho- \tilde T_0 \tilde
  \eta +\tilde  P_0,
  \end{equation}
  or equivalently, by using   \eqref{specialvolumeenergy},
\begin{equation}
   (2 \tilde\Lambda-1)\, \tilde \varepsilon=2\tilde\Lambda\,  \tilde \varepsilon_0-\tilde \mu_{_0}\tilde \rho- \tilde T_0 \tilde
  \eta +\tilde  P_0 , \label{IP2}
  \end{equation}
  where the constant $\tilde P_0$ has the dimension of a pressure.

\medskip In the bulk  the fields become constant and
the equilibrium equations lead to
$$\tilde\Lambda\tilde \mu-  \tilde \mu_{_0}=0,\quad
\tilde\Lambda \tilde  T-\tilde T_0  =0 ,\quad
1- \tilde\Lambda=0, \quad \tilde \varepsilon-\tilde \mu_{_0}\tilde \rho- \tilde T_0 \tilde
  \eta +\tilde  P_0=0.$$
Hence $\tilde \Lambda=1$ and $\tilde \mu_{_0}$,  $\tilde T_0$,
$\tilde P_0$ are respectively the common values of the chemical
potential, temperature and pressure in both bulk phases and  we
recover the usual global equilibrium conditions for planar
interfaces.

We denote by superscripts ${}^+$ and ${}^-$ the values of the
fields in the two bulk phases. From \eqref{Lzero}, \eqref{temperature},  \eqref{chempote} we
deduce the equalities of thermodynamical quantities $\tilde\mu_0,
\tilde T_0, \tilde P_0$ in the two bulks phases
\begin{align}
  2 \tilde \eta^+\tilde \rho^++  4 (\tilde \rho^+)^3&= 2 \tilde \eta^-\tilde \rho+
  4 (\tilde \rho^-)^3=\tilde\mu_0\label{eqmu}\\
  2 \tilde\chi \tilde \eta^+ +  (\tilde \rho^+)^2&=2 \tilde\chi \tilde
  \eta^- +  (\tilde \rho^-)^2=\tilde T_0\label{eqT}\\
   \tilde\chi (\tilde \eta^+)^2+2 \tilde \eta^+ (\tilde \rho^+)^2
   + 3 (\tilde \rho^+)^4&=\tilde\chi (\tilde \eta^-)^2+2 \tilde \eta^-
   (\tilde \rho^-)^2 + 3 (\tilde \rho^-)^4=\tilde P_0.  \label{eqP}
\end{align}
Using (\ref{eqT}), equations (\ref{eqmu}) and (\ref{eqP}) can be
written
\begin{align*}
 \tilde T_0 \tilde \rho^+ + (4\tilde\chi-1)(\tilde \rho^+)^3&=
 \tilde T_0 \tilde \rho^- + (4\tilde\chi-1)(\tilde \rho^-)^3=\tilde\chi \tilde
 \mu_0 ,\\
  2\tilde T_0 (\tilde \rho^+)^2 + 3(4\tilde\chi-1) (\tilde \rho^+)^4&= 2\tilde T_0 (\tilde \rho^-)^2 + 3(4\tilde\chi-1) (\tilde \rho^-)^4=4\tilde\chi\tilde P_0- \tilde T_0^2 ,
\end{align*}
which implies
\begin{align*}
  &\Big(\tilde T_0  + (4\tilde\chi-1)\big((\tilde \rho^+)^2+\tilde \rho^+ \tilde \rho^-+ (\tilde \rho^-)^2\big)
   \Big) \big(\tilde \rho^+- \tilde \rho^-\big)= 0,\\
   &\Big(2\tilde T_0 + 3(4\tilde\chi-1) \big((\tilde \rho^+)^2+ (\tilde \rho^-)^2\big)\Big)
   \big((\tilde \rho^+)^2- (\tilde \rho^-)^2\big)= 0.
\end{align*}
 Considering
an interface between two distinct phases, we have $\tilde
\rho^+\not= \tilde \rho^-$, thus
\begin{align*}
  &\tilde T_0  + (4\tilde\chi-1)\big((\tilde \rho^+)^2+\tilde \rho^+ \tilde \rho^-+ (\tilde \rho^-)^2\big) = 0,\\
  &\Big(2\tilde T_0 + 3(4\tilde\chi-1) \big((\tilde \rho^+)^2+ (\tilde \rho^-)^2\big)\Big) \big(\tilde \rho^++ \tilde \rho^-\big)= 0.
\end{align*}
Subtracting to the second equation the product of the first one
by $2 \big(\tilde \rho^++ \tilde \rho^-\big)$ the system becomes
\begin{align*}
  & (4\tilde\chi-1) \big(\tilde \rho^+- \tilde \rho^-\big)^2 \big(\tilde \rho^+ + \tilde \rho^-\big)= 0 \\
  &\tilde T_0  + (4\tilde\chi-1)\big((\tilde \rho^+)^2+\tilde \rho^+ \tilde
  \rho^-+ (\tilde \rho^-)^2\big) = 0.
\end{align*}
As expected this system admits no solution when $\tilde T_0>0$, or equivalently when the temperature in the phases is greater than the critical one.
Let us set $\tilde T_0:=-(4\tilde\chi-1)\, \kappa^2$, i.e.
\begin{equation}
  \kappa:=\sqrt{\frac {-\tilde T_0}{4\tilde\chi-1}}. \label{estimation0}
\end{equation}
The  small quantity
$\kappa$ measures the distance  from the critical point. Using again $\tilde \rho^+\not= \tilde \rho^-$ we
find
\begin{equation}
 \tilde \rho^+=\kappa\quad {\rm and}\quad \tilde \rho^-=-\kappa, \label{estimation1}
\end{equation}
from which we directly deduce,
\begin{equation}
\tilde \eta^+=  \tilde \eta^-=-2\,\kappa^2 ,\quad \tilde
\mu_0=0,\quad \tilde \varepsilon^+=  \tilde  \varepsilon^-=\tilde P_0=(4\chi-1)\kappa^4 .
\label{estimation2}
\end{equation}

\subsection{The rescaling process}

 {In view of Eqs. \eqref{estimation0}, \eqref{estimation1}, \eqref{estimation2} the values of $\tilde \rho$ and $\tilde \eta$ in the
phases lead to the natural rescaling
\begin{equation}
\check \rho:= \kappa^{-1} \tilde\rho,\quad \check \eta :=
\kappa^{-2} \tilde\eta,\quad \check \varepsilon:= \kappa^{-4} \tilde\varepsilon, \quad\check z:= \kappa z\label{rescal1}
\end{equation}
 and system \eqref{systemA} becomes
\begin{equation}
\left\{
\begin{array}{l}
 \tilde C_0\,(\tilde \Lambda \check \rho')' + \tilde D_0\,\kappa\, (\tilde \Lambda\,\check \eta')'+\tilde F_0\, \kappa^3\,(\tilde \Lambda\check  \varepsilon')' =\tilde \Lambda \big(2  \check \eta\check \rho+  4 \check \rho^3\big),    \\
 \tilde D_0\,\kappa\, (\tilde \Lambda\,\check  \rho')' + \tilde E_0\,\kappa^2\,(\tilde \Lambda\check
\eta')' +\tilde G_0\, \kappa^4\,(\tilde \Lambda\check \varepsilon')'=
 \tilde \Lambda \big( 2 \tilde\chi \check \eta+ \check  \rho^2 \big)+  (4\tilde\chi-1), \\
 \tilde   F_0\, \kappa^3\, (\tilde \Lambda\check \rho')' + \tilde G_0\,\kappa^4\, (\tilde \Lambda\check
\eta')' +\tilde H_0\, \kappa^6\, (\tilde \Lambda\check\varepsilon')'=  1- \tilde
\Lambda , \label{systemC}
\end{array}%
\right.
\end{equation}
where the space derivatives are now relative to $\check z$. Hence $\tilde\Lambda=1+O(\kappa^3)$ and
at the first order with respect to the small parameter
$\kappa$,
\begin{equation}
\left\{
\begin{array}{l}
\displaystyle  \tilde C_0\, \check \rho''=2 \check \eta\check \rho+  4 \check \rho^3,    \\
\displaystyle 0 = 2  \tilde\chi \check \eta + \check \rho^2
+(4\tilde\chi-1),
\end{array}%
\right.
\label{doubleetoile}\end{equation}
which gives  by elimination of $\check \eta$,
\begin{align}
  &\frac{\tilde\chi\tilde C_0 }{(4\tilde\chi-1)}\, \check \rho''=\check \rho^3 - \check \rho.\label{eqdif}
\end{align}
Multiplying by $\check \rho'$, integrating and taking into account \eqref{estimation1}, we get
\begin{align}
  &\frac{\tilde\chi\tilde C_0 }{(4\tilde\chi-1)}\,\frac { \check \rho'^2} 2=\frac 1 4 \left(\check \rho^2 - 1 \right)^2.\label{IPrho}
\end{align}
Hence the mass density profile $\check \rho_{eq}$ at equilibrium  across an interface has the classical representation (cf. \cite{Rowlinson} p. 251)}
\begin{equation}
\check \rho_{eq} (\check z)=\tanh(\frac {\check z}{\ell})
\label{densityprofile}
\end{equation}
where
\begin{equation}
\ell=\sqrt{\frac{2 \tilde\chi\check C_0
}{(4\tilde\chi-1)}}.\label{ell}
\end{equation}
Note that this well known profile is an exact solution of \eqref{eqdif} but results from several approximations. It is valid only for a planar interface lying far from the boundaries of the domain.  Moreover considering the polynomial form \eqref{Lzero} for the energy is clearly an approximation as well as neglecting the terms of lower order in \eqref{eqmu}, \eqref{eqT} and \eqref{eqP}.

Using \eqref{temperature} and \eqref{doubleetoile} we obtain that the temperature through the interface is constant at the first order:
$$\check T_{eq}= 2  \tilde\chi \check \eta_{eq} + (\check
\rho_{eq})^2 =-(4\tilde\chi-1) .
$$
\begin{figure}[h]
\begin{center}
\includegraphics[width=8cm]{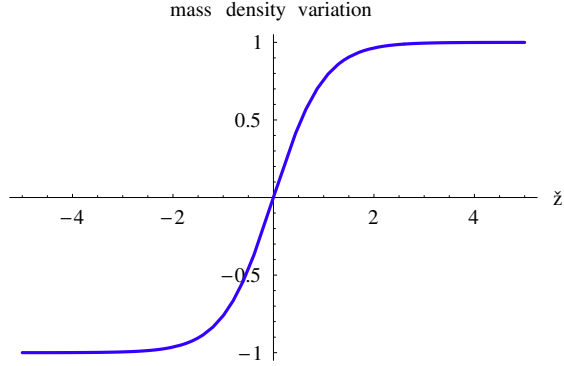}
\end{center}
\caption {Classical density profile of the
normalized density (\,$\check \rho \in ]-1,+1[$\,) associated with
\eqref{densityprofile} and \eqref{ell}; the $ x$-axis unit is
$\ell$.} \label{fig2}
\end{figure}
However the second equation of system \eqref{systemC} gives a more
accurate information about the temperature profile through the
interface; indeed, at  order $\kappa$,
\begin{equation}
\tilde D_0\,  \kappa\, \check \rho'' = 2  \tilde\chi \check \eta + \check \rho^2
+(4\tilde\chi-1)+ O(\kappa^2). \label{system1}
\end{equation}
That is
\begin{align*}
  \check  T_{eq}&=-(4\tilde\chi-1)+  \kappa\, \tilde D_0\, {\check\rho_{eq}''}+O(\kappa^2)\\
  &=-(4\tilde\chi-1)+\kappa\, \frac {(4\tilde\chi-1)}{\tilde\chi} \, \frac {\tilde D_0} {\tilde C_0}
  \,\left[  {\check\rho_{eq}^3}- {\check\rho_{eq}}\right]+O(\kappa^2).
\end{align*}
Consequently,
\begin{equation}
\check T_{eq} = (4\tilde\chi-1)\Bigg( -1+
\frac{\kappa}{\tilde\chi} \,  \frac {\tilde D_0}{\tilde C_0} \left( {\tanh^3\left(\frac {\check z}{\ell}\right)} -
{\tanh\left(\frac {\check z}{\ell}\right)}\right)\Bigg)+O(\kappa^2).
\label{temprofile}
\end{equation}
\begin{figure}[h]
\begin{center}
\includegraphics[width=9cm]{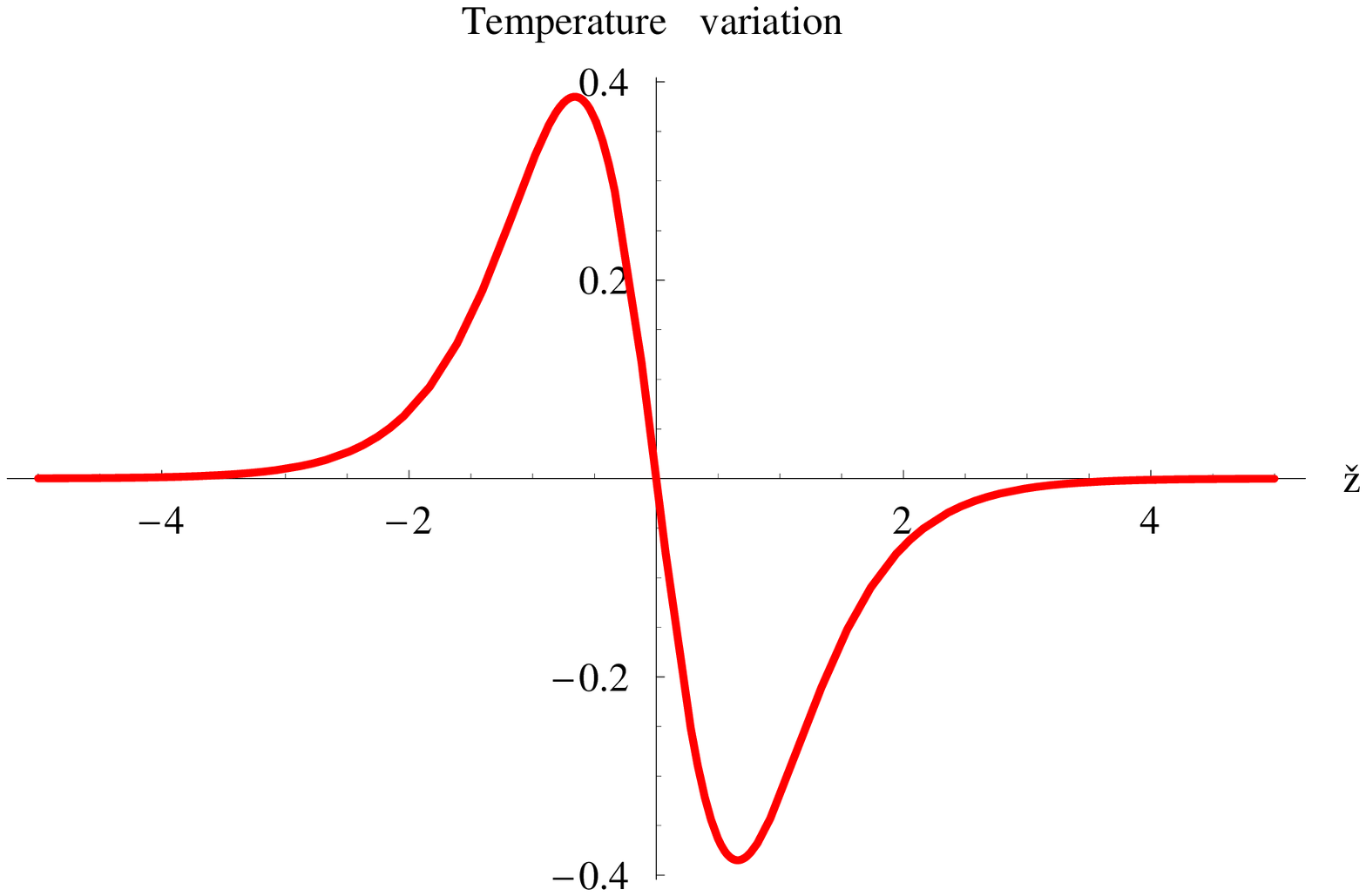}
\end{center}
\caption{Variation of  normalized temperature $\check T_{eq}+(4\chi-1)$
 through the interface near the critical point.
The $x$-axis unit is $\ell$ and the $y$-axis unit is  $\displaystyle \kappa\,
\frac{(4\tilde\chi-1)}{\tilde\chi} \,  \frac {\tilde D_0}{\tilde
C_0}$. \label{fig3}}
\end{figure}
Note that in Eq.\eqref{temprofile} the variation of the temperature across the interface is no more monotonic (see Fig. \ref{fig3}). Moreover, the variation of temperature $\check T_{eq}$ is multiplied by the small parameter $\kappa$ and    is negligible with respect to the variation of $\check \rho_{eq}$.

\section{Surface tension}
        Surface tension $\sigma $ of a plane liquid-vapor interface corresponds to the excess of  free energy $  \tilde e:=  \tilde \varepsilon-  \tilde T   \tilde \eta$ inside the interface.  Using \eqref{temperature} and \eqref{energyinterne}, we have
            $$   \tilde e=\frac {4\tilde \chi-1} {4\tilde \chi} \big( \tilde \rho^4-2\tilde\kappa^2   \tilde \rho^2-\kappa^4 (4\tilde \chi-1)  \big) +\tilde Q( \tilde \rho',\tilde \eta', \tilde \varepsilon') $$
            As, in the bulk, we have
            $$   \tilde  e^+=  \tilde e^-=-(4 \tilde \chi-1)\kappa^4,  $$ surface tension is
 \begin{align}
   \tilde\sigma &:=\int_{-\infty}^{+\infty}  \left( \tilde  e+ (4 \tilde \chi-1)\kappa^4\right) dz
   =\int_{-\infty}^{+\infty}  \left(\frac {4\tilde \chi-1} {4\tilde \chi}( \tilde \rho^2-\tilde\kappa^2)^2   +\tilde Q( \tilde \rho',\tilde \eta', \tilde \varepsilon')\right) dz\nonumber \\
   &=\kappa^3 \int_{-\infty}^{+\infty}  \left(\frac {4\tilde \chi-1} {4\tilde \chi}( \check \rho^2-1)^2   +\tilde Q(\check \rho',\kappa \check \eta',\kappa^3 \check \varepsilon')\right) d\check z .
\end{align}
 At the first order with respect to $\kappa$, we obtain
 \begin{align} \tilde\sigma &=\kappa^3 \int_{-\infty}^{+\infty}  \left(\frac {4\tilde \chi-1} {4\tilde \chi}( \check \rho^2-1)^2   +\frac 1 2 \tilde C_0\,  \check \rho'^2\right) d\check z +O(\kappa^4) \label{34} \\
   &= \kappa^3 \int_{-1}^{+1}  \left(\sqrt{
   \frac {(4\tilde \chi-1) \tilde C_0} {2\tilde \chi}}\, (1- \check \rho^2)
   \right) d\check  \rho +O(\kappa^4)\nonumber \\
    &= \kappa^3 \, \frac 4 3 \,  \sqrt{
   \frac {(4\tilde \chi-1) \tilde C_0} {2\tilde \chi}}\,  +O(\kappa^4)
 \end{align}
 Thus, at the leading order, equilibrium values and surface tension are those given by the Cahn-Hilliard theory : the effect of the gradients of entropy and energy densities are negligible. A more accurate description could be obtained : terms of order $\kappa^4$ would come from \emph{(i)} the perturbation of system \eqref{doubleetoile} by taking into account the coupling term $\tilde D_0$ and \emph{(ii)} the introduction of the same coupling term in \eqref{34}.\\

\section{Conclusion}

We have obtained the mass density and temperature profiles through
an interface near the critical point. Our results present some similarities with
the ones  obtained in \cite{Ruggeri} for fluid mixtures where two
mass densities have the role played here by mass and entropy
densities. The differences lie  in the fact that we are not here
impelled to deal with combinations of densities and also in the fact that the notion of critical point is more complex in the case of a mixture where non-monotonic profiles can be obtained at the leading order.

We have
introduced a state law in which all gradients are considered with
respect to mass, entropy and \emph{energy} densities. At our
knowledge, it is the first time that this though natural
assumption is used. In this framework, we confirm the conjecture
made by Rowlinson and Widom \cite{Rowlinson} that, near the
critical point, the variations of temperature inside the
interfacial layer are negligible. This result is mainly due to the
fact that the variations of entropy  density are negligible with
respect to the variations of mass density.\\

\textbf{Data accessibility statement}. This work does not have any experimental data.\\

\textbf{Competing interests statement}. We have no competing interest. \\

\textbf{Author's contribution}. H.G. and P.S. conceived the mathematical model, interpreted the results, and wrote together the paper.\\

\textbf{Acknowledgements}. P.S. thanks the Laboratoire de M\'ecanique et d'Acoustique (Marseille) for its hospitality.\\

\textbf{Funding statement.} This work was supported by C.N.R.S.\\

\textbf{Ethics statement}. This work is original, has not previously been published in any Journal and is not currently under consideration for publication elsewhere.\\

\end{document}